\input harvmac


%
%
\ifx\epsfbox\UnDeFiNeD\message{(NO epsf.tex, FIGURES WILL BE IGNORED)}
\def\figin#1{\vskip2in}
\else\message{(FIGURES WILL BE INCLUDED)}\def\figin#1{#1}\fi
\def\ifig#1#2#3{\xdef#1{fig.~\the\figno}
\goodbreak\midinsert\figin{\centerline{#3}}%
\smallskip\centerline{\vbox{\baselineskip12pt
\advance\hsize by -1truein\noindent\footnotefont{\bf Fig.~\the\figno:}
#2}}
\bigskip\endinsert\global\advance\figno by1}

\def\ifigure#1#2#3#4{
\midinsert
\vbox to #4truein{\ifx\figflag\figI
\vfil\centerline{\epsfysize=#4truein\epsfbox{#3}}\fi}
\narrower\narrower\noindent{\footnotefont
{\bf #1:}  #2\par}
\endinsert
}


\def\IC{{\ \hbox{{\rm I}\kern-.6em\hbox{\bf C}}}}
\def\IR{{\hbox{{\rm I}\kern-.2em\hbox{\rm R}}}}
\def\IZ{{\hbox{{\rm Z}\kern-.4em\hbox{\rm Z}}}}
\def\pa{\partial}
\def\sIR{{\hbox{{\sevenrm I}\kern-.2em\hbox{\sevenrm R}}}}

\def\dzb{D0 branes}
\def\mt{M-theory}
\def\iml{infinite momentum limit}
\def\imf{infinite momentum frame}
\def\lm{longitudinal momentum}
\def\sg{supergravity}
\def\p{p_{11}}
\def\x{x^{11}}

\def\yr{Y_{rel}}

\Title{RU-96-95, SU-ITP-96-12, UTTG - 13 -96}{\vbox{\centerline{M Theory
As A Matrix Model:}\centerline{ A Conjecture}}}

\centerline{\it T. Banks~$^1$, W. Fischler~$^2$,  S.H. Shenker~$^1$, 
L. Susskind~$^3$}
\medskip
\centerline{$^1$Department of Physics and Astronomy}
\centerline{Rutgers University, Piscataway, NJ 08855-0849}
\centerline{\tt banks, shenker@physics.rutgers.edu}
\medskip
\centerline{$^2$Theory Group, Department of Physics}
\centerline{University of Texas, Austin, TX, 78712}
\centerline{\tt fischler@physics.utexas.edu}
\medskip
\centerline{$^3$Department of Physics}
\centerline{Stanford University, Stanford, CA 94305-4060}
\centerline{\tt susskind@dormouse.stanford.edu}

\bigskip

\medskip

\noindent

We suggest and motivate a precise equivalence between uncompactified
eleven dimensional \mt \ and the $ N=\infty$ limit of the supersymmetric matrix
quantum mechanics describing \dzb \ . The evidence for the conjecture
consists of several correspondences between the two theories. As a
consequence of supersymmetry the simple matrix model is rich enough to describe
the properties of the entire Fock space of massless well separated particles of
the supergravity theory. In one particular kinematic situation the leading
large distance interaction of these particles is 
exactly described by supergravity .
The model appears to be a nonperturbative realization of the holographic
principle. The membrane states required by \mt \ are
contained as excitations of the matrix model. The membrane world  volume is a
noncommutative geometry embedded in a noncommutative spacetime.

\hyphenation{Min-kow-ski}
\Date{October 1996}


%
%

%
%
\newsec{\bf Introduction}

M theory \ref\dthw{M.J. Duff, P. Howe,
T. Inami, K.S. Stelle, "Superstrings in D=10 
{}from supermembranes in D=11", Phys. Lett. B191 (1987) 70\semi
M.J. Duff, J. X. Lu, "Duality rotations in membrane theory", Nucl. 
Phys. B347 (1990) 394 \semi M.J. Duff, R. Minasian, James T. Liu, "Eleven-dimensional origin of 
string/string duality: a one-loop test", Nucl. Phys. B452 (1995) 261 \semi
C. Hull and P. K. Townsend, "Unity of superstring dualities",
Nucl. Phys. { B438} (1995) 109,
hep-th/9410167
\semi
P. K. Townsend, ``The Eleven-Dimensional Supermembrane Revisited''
 Phys. Lett. B350 (1995) 184 hep-th/9501068 \semi
"String-membrane duality in seven dimensions", Phys. Lett. { 354B} 
(1995) 247, hep-th/9504095 \semi
C. Hull and P. K. Townsend,
``Enhanced gauge symmetries in superstring theories", Nucl. Phys. { B451} 
(1995) 525, hep-th/9505073\semi
E. Witten ``String Theory Dynamics in Various Dimensions",
Nucl.Phys. B443 (1995) 85 hep-th/9503124. }
is the strongly coupled limit of type IIA
string
theory. In the limit of infinite coupling it becomes an eleven dimensional
theory in a background infinite flat space. In this paper \mt \ will always
refer to this decompactified limit.   We know very little about this theory
except for the following two facts. At low energy and large distances it is
described by eleven dimensional \sg \ . It is also known to possess membrane
degrees of freedom with membrane tension $1\over l_p^3$ where
$l_p$ is the eleven dimensional Planck length. It seems extremely unlikely that
\mt \ is any kind of conventional quantum field theory. The degrees of freedom
describing the short distance behavior are simply unknown. The purpose of this
paper is to put forward conjecture about these degrees of
freedom and about the Hamiltonian governing them.

The conjecture grew out of a number of disparate facts about \mt \ ,
 D~branes~\ref\jopo{J.Polchinski, ``Dirichlet Branes and Ramond-Ramond
charges",  Phys.Rev.Lett.75 (1995) 4724\semi for a review see J.Polchinski,
S.Chaudhuri, and C.V. Johnson, ``Notes on D-Branes", ITP preprint
NSF-ITP-96-003, hep-th/9602052 .}, matrix descriptions of their dynamics
\ref\w{E.Witten, ``Bound States of Strings and p-Branes Nucl.Phys. B 460, 1995,
335 hep-th/9510135. }, supermembranes \nref\ghdnb{E. Bergshoeff, E. Sezgin and
P.K. Townsend Phys.Lett. B189 (1987) 75; Ann. Phys. (NY) 185 (1988) 330\semi
P.K. Townsend in ``Superstring '88", proc.of the Trieste Spring School, eds
M.B. Green, M.T. Grisaru, R. Iengo and A. Strominger ( World Scient., 1989
)\semi M.J. Duff, Class. Quantum Grav. 5 (1988) 189}\nref\hoppe{B. de Wit, J.
Hoppe and H. Nicolai, Nucl.Phys. B 305 [FS 23] (1988)
545}\nref\luscher{B. de Wit, M.
Luscher and H. Nicolai, Nucl. Phys. B 320 (1989) 135}
\refs{\ghdnb,\hoppe,\luscher}, the holographic principle
\ref\holo{G.'t Hooft, ``Dimensional Reduction in Quantum Gravity" Utrecht
preprint THU-93/26 gr-qc/9310026\semi L.Susskind, ``The World as a Hologram" J.
Math. Phys. 36 (1995) 6377, hep-th/940989.}, and
\nref\short{S. H. Shenker,
``Another Length Scale in String Theory?'', hep-th/9509132.}
\nref\zerobranes{U.H. Danielsson, G. Ferretti and B. Sundborg,``D-particle
Dynamics and Bound States", hep-th/9603081\semi D. Kabat and P. Pouliot,``A
Comment on Zero-Brane Quantum Mechanics", hep-th/9603127\semi M.R. Douglas, D.
Kabat, P. Pouliot and S. Shenker, ``D-Branes and Short Distance in String
Theory", hep-th/9608024 .}
short distance phenomena in string theory \refs{\short, \zerobranes}. 
Simply stated the  conjecture is
this. \mt\ , in the light cone frame, is exactly described by the 
large $N$ limit of a
particular supersymmetric matrix quantum mechanics. The system is the same one
that has been used previously used to study the small distance behavior
of \dzb\ \zerobranes .
P.K. Townsend~\ref\townsrem{P.K. Townsend, 
"D-branes {}from M-branes", Phys. Lett. { B373} (1996) 68,
hep-th/9512062.}\ was the first to point out that the supermatrix
formulation of membrane theory suggested that membranes could be viewed
as composites of \dzb  .  Our work is a precise realization of his
suggestion. 

In what follows we will present our conjecture and some evidence for it.
We begin by reviewing the description of string theory in the \imf .  We
then present our conjecture for the full set of degrees of freedom of
\mt \ and the Hamiltonian which governs them.  Our strongest evidence for
the conjecture is a demonstration that our model contains the
excitations which are widely believed to exist in \mt \ , supergravitons
and large metastable classical membranes.  These are discussed in
sections $3$ and $5$.  The way in which these excitations arise is
somewhat miraculous, and we consider this to be the core
evidence for our conjecture.
In section $4$ we present a calculation of
supergraviton scattering in a very special kinematic region, and argue
that our model reproduces the expected result of low energy
supergravity.  The calculation depends on a supersymmetric
nonrenormalization theorem whose validity we will discuss there.  In
section $6$ we argue that our model may satisfy the Holographic
Principle.  This raises crucial issues about Lorentz invariance which are
discussed there.

We emphasize that there are many unanswered questions about our proposed
version of \mt .  Nonetheless these ideas seem of sufficient interest
to warrant presenting them here.
If our conjecture is correct, this would be
the first nonperturbative formulation of a quantum theory which includes
gravity.   

%
%
\newsec{The \imf \ and the Holographic Principle}

The \imf\ \ref\imflit{S. Weinberg,``Dynamics at Infinite Momentum", Phys.Rev.
150 (1966) 1313\semi J. Kogut and L.Susskind, ``The Parton Picture of Elementary
Particles", Physics Reports 8 (1973) 75.} is the old name for the  misnamed
light cone frame. Thus far this is the only frame in which it has proved
possible to formulate  string theory in Hamiltonian form. The
description  of \mt \
which we will give in this paper is also in the \imf \ . We will begin by
reviewing some of the features of the \imf \ formulation of relativistic
quantum
mechanics. For a comprehensive review we refer the reader to \imflit . We
begin
by
choosing a particular spatial direction
$x^{11}$ called the longitudinal direction. The nine dimensional space
transverse to
$x^{11}$ is labeled $x^i$ or $x^{\perp}$. Time will be indicated by $t$. Now
consider a system of particles with momenta $(p^a_{\perp}, p^a_{11})$ where $a$
labels the particle. The system is boosted along the $x^{11}$ axis until all
longitudinal momenta are much larger than any scale in the problem. Further
longitudinal boosting just rescales all longitudinal momenta in fixed
proportion. Quantum field theory in such a limiting
reference frame has a number of properties which will be relevant to us.

It is convenient to begin by assuming the $\x$ direction is compact with a
radius $R$. The compactification serves as an infrared cutoff. Accordingly,
the
longitudinal
momentum of any system or subsystem of quanta is quantized  in units
of  $1/R$. In the \imf \ all systems are composed of constituent quanta or
partons. The partons all carry strictly positive values of longitudinal
momentum.
It is particularly important to understand what happens to quanta of negative
or vanishing $\p$. The answer is that as the \iml \ is approached the frequency
of these quanta, relative to the Lorentz time dilated motion of the boosted
system becomes infinite and the zero and negative momentum quanta may be
integrated out. The process of integrating out such fast modes may influence or
even determine the Hamiltonian of the remaining modes. In fact the situation is
slightly more complicated in certain cases for the zero momentum degrees of
freedom. In certain situations such as spontaneous symmetry breaking, these
longitudinally homogeneous modes define backgrounds whose moduli may appear in
the Hamiltonian of the other modes. In any case the zero and negative momentum
modes do not appear as independent dynamical degrees of freeedom.

Thus we may assume all systems have longitudinal momentum given by an  integer
multiple of $1/R$
\eqn\ppar{
p_{11} = N/R}
with N strictly positive. At the end of a calculation we must let $R$ and $N/R$
tend to infinity to get to the uncompactified  infinite momentum limit.

The main reason for the simplifying features of the \imf \ is the existence of
a transverse Galilean symmetry which leads to a naive nonrelativistic
form for the equations. The role of nonrelativistic mass is played by the
longitudinal momentum $p_{11}$. The Galilean transformations take the form
\eqn\galilei{
p_i \to p_i + p_{11}\ v_i}

As an example of the Galilean structure of the equations, the energy of a
free massless particle is
\eqn\emasslss{
E={ p_{\perp}^2\over 2p_{11}}}

For the eleven dimensional supersymmetric theory we will
consider, the Galilean invariance is extended to the Supergalilean group
which includes 32 real supergenerators. The supergenerators divide into two
groups of 16, each transforming as spinors under the nine dimensional 
transverse rotation group. We denote them by $Q_{\alpha} $ and $q_{A}$ 
and they obey anticommutation relations
\eqn\supalg{
\eqalign{[Q_{\alpha},Q_{\beta}]_+ &= \delta_{\alpha\beta} H \cr
         [q_A,q_B]_+ &= \delta_{AB}P_{11} \cr
         [Q_{\alpha} ,q_A ] &= \gamma^i_{A\alpha}  P_i \cr}}
The Lorentz generators which do
not preserve the \imf \ mix up the two kinds of generators. 

Let us now recall some of the features of string theory in the infinite
momentum or light cone frame \ref\thorn{C.B. Thorn, Proceedings of
Sakharov Conf. 
on Physics Moscow (1991) 447-454, hep-th/9405069.}.
We will continue   to call the
longitudinal direction $x^{11}$ even though in this case the theory has only
ten
space-time directions. The transverse space is of course eight dimensional. To
describe a free string of \lm \ $p_{11}$ a periodic parameter  $\sigma$ which
runs {}from
$0$ to $p_{11}$ is introduced. To regulate the world sheet theory a cutoff
$\delta \sigma = \epsilon$ is introduced. This divides the parameter space into
$N=\p /{\epsilon}$ segments, each carrying \lm \ $\epsilon$. We may think of
each segment as a parton but unlike the partons of quantum field theory these
objects always carry $\p = \epsilon$. For a multi particle system of total \lm
\ $\p (total)$ we introduce a total parameter space of overall length $\p
(total)$ which we allow to be divided into separate pieces, each describing a
string. The world sheet regulator is implemented by requiring each string to be
composed of an integer number of partons of momentum $\epsilon$. Interactions
are described by splitting and joining processes in which the number of partons
is strictly conserved. The regulated theory is thus seen to be a special case
of Galilean quantum mechanics of $N$ partons with interactions which bind them
into long chains and allow particular kinds of rearrangements.

 The introduction of a minimum unit of momentum $\epsilon$
can be given an interpretation as an infrared cutoff. In particular we may
assume
that the
$x^{11}$ coordinate is periodic with length $R= {\epsilon ^{-1}}$. Evidently
the physical limit
$\epsilon
\to
0, R\to \infty$ is a limit in which the number of partons
$N$ tends to infinity.

It is well known \holo\ that in this large $N$ limit the partons become
infinitely dense in the transverse space and that this leads to extremely
strong
interactions. This circumstance together with the Bekenstein bound on entropy,
has led to the  $holographic$ speculation that the transverse density of
partons is strictly bounded to about one per transverse Planck area. In other
words the partons form a kind of incompressible fluid.  This leads to the
unusual consequence that the transverse area occupied by a system of \lm
\
$\p
$ can not be smaller than $\p / \epsilon$ in Planck units.

The general arguments for the holographic behavior of systems followed from
considerations involving the Bekenstein - 't Hooft bound on the entropy of a
spatial region~\ref\thooft{J. D. Bekenstein, Phys.Rev. D 49 (1994), 6606;
G. 't Hooft in ref \holo.} 
and were not specific to string
theory. If the arguments are correct they should also apply to 11 dimensional
theories which include gravitation. Thus we should expect that in M-theory the
radius of a particle such as the graviton will grow with
$\p$ according to
\eqn\size{
\rho = ({\p \over \epsilon})^{1\over 9}{l_p} = (\p R)^{1\over 9}{l_p}}
where $l_p$ is the eleven dimensional Planck length. In what follows we will
see quantitative evidence for exactly this behavior.

At first sight the holographic growth of particles appears to contradict the
boost invariance of particle interactions. Consider the  situation of two low
energy particles moving past one another with some large transverse separation,
let us say of order a meter. Obviously these particles have negligible
interactions. Now boost the system along the longitudinal direction until the
size of each particle exceeds their separation. They now overlap as
they pass each other. But longitudinal boost invariance requires that the
scattering amplitude is still essentially zero. This would seem to require
extremely special and unnatural cancellations. We will see below that
one key to this behavior is the
very special BPS property of the partons describing M-theory.  However,
we are far from having a complete understanding of the longitudinal
boost invariance of our system.  Indeed, we view it as the key dynamical
puzzle which must be unravelled in understanding the dynamics of \mt .

%
%
\newsec{M-theory and \dzb }

M-theory with a compactified longitudinal coordinate $\x$ is by definition type
IIA string theory. The correspondences between the two theories include~
 \dthw\ :

\item{1.}
The compactification radius $R$ is related to the string coupling constant
by
\eqn\Relevg{R=g^{2/3}l_p =g l_s}

where $l_s$ is the string length scale
\eqn\lelevls{l_s = g^{-1/3}l_p}

\item{2.}
The Ramond-Ramond photon of IIA theory is the Kaluza Klein photon which arises
upon compactification of eleven dimensional supergravity.

\item{3.}
No perturbative string states carry  RR charge. In other words all
perturbative string states carry vanishing momentum along the $\x$
direction. The only objects in the theory which do carry RR photon
charge are the
\dzb \ of Polchinski. \dzb \ are point particles which carry a single
unit of RR charge. Equivalently they carry \lm \
\eqn\zerobranemom{
P_{11}^{D0\ branes }  = 1/R}
The \dzb \ carry the quantum numbers of the first massive KK modes of the
basic
eleven dimensional
supergravity multiplet, including 44 gravitons, 84 components of a 3-form, and
128 gravitinos. We will refer to these particles as supergravitons.  As
11-dimensional objects these are all massless. As a consequence they are BPS
saturated states in the 10-dimensional theory. Their 10-D mass is $1/R$.

\item{4.}
Supergravitons carrying Kaluza Klein momentum $\p = N/R$ also exist but are
not described as elementary \dzb. As shown in \w \ their proper description
is as bound composites of $N$ \dzb.

These properties make the \dzb \ candidate partons for an \iml
\
description of M-theory. We expect that if, as in quantum field theory, the
degrees of freedom with vanishing and negative $\p$ decouple then M-theory in
the \imf
\ should be a theory whose only degrees of freedom are \dzb. Anti-\dzb \ carry
negative Kaluza Klein momenta and strings carry vanishing $\p$. The decoupling
of anti-\dzb \ is particularly fortunate because brane anti-brane dynamics is
something about which we know very little \ref\antibrane{T. Banks and
L.Susskind, ``Brane - Antibrane Forces", hep-th/9511194.}.  The BPS property of
zero branes ameliorates the conflict between infinitely growing
parton wave functions and low energy locality which we noted at the end of the
last section.  We will 
see some partial evidence for this in a nontrivial scattering
computation below.
We will also discuss below the important point that a model
containing only \dzb \ actually
contains large classical supermembrane excitations.  Since the
conventional story of the M theoretic origin of strings depicts them as
membranes wrapped around the compactified eleventh dimension, we have
some reason to believe that strings have not really been left out of the
system.

All of these
circumstances lead us to propose that M-theory in the \imf \ is a theory in
which the only dynamical degrees of freedom or partons are \dzb. Furthermore it
is clear in this case that all systems are built out of composites of partons,
each of which carries the minimal $\p$.  We note however that our system
does have a set of degrees of freedom which go beyond the parton
coordinates.  Indeed, as first advocated in \w , the $D0$ brane coordinates of
$N$ partons have to be promoted to matrices.  At distance scales larger
than the eleven dimensional Planck scale, these degrees of freedom
become very massive and largely decouple\foot{Indeed
 we will propose that this decoupling is precisely
what defines the regime in which the classical notion of distance makes
sense.}, but their virtual effects are
responsible for all parton interactions.  These degrees of freedom are
BPS states, and are related to the parton coordinates by gauge
transformations.  Furthermore, when the partons are close together they
become low frequency modes.  Thus they cannot be omitted in
any discussion of the dynamics of \dzb \ .

%
%
\newsec{\bf D0 brane Mechanics}

If the \iml \ of M-theory is the theory of \dzb, decoupled from the other
string theory degrees of freedom, what is the precise form of the
quantum mechanics of the system? Fortunately there is a very good candidate
which has been extensively studied in another context in which \dzb \ decouple
{}from strings \zerobranes .

As emphasized at the end of the last section, open strings which 
connect \dzb \ do not exactly decouple. In fact the very short strings which
connect the branes when they are practically on top of each other introduce a
new kind of coordinate space in which the nine spatial coordinates of a
system of N \dzb \ become   nine   $N\times N$ matrices, $X^i_{a,b}$ \w.
The matrices $X$ are accompanied by 16 fermionic superpartners $\theta_{a,b}$
which transform as spinors under the $SO(9)$ group of transverse rotations.
The matrices may be thought of as the spatial components of 10 dimensional
Super
Yang Mills (SYM) fields after dimensional reduction to zero space directions.
These Yang Mills fields describe the open strings which are attached to the
\dzb. The Yang Mills quantum mechanics has $U(N)$ symmetry and is described (in
units with $l_s=1$) by the Lagrangian
\eqn\lag{L={1\over 2g}\biggl[ \tr \dot{ X^i}\dot{ X^i} + 2{\theta^T}
 \dot{\theta} - {1\over 2}\tr[X^i,X^j]^2 - 
2{\theta^T} \gamma_i [\theta ,X^i ]
\biggr]}
Here we have used conventions in which the fermionic variables are 
$16$ component nine dimensional spinors.

In \zerobranes\ this Lagrangian was used to
to study the short distance properties of \dzb \ in weakly
coupled string theory.  The 11-D Planck length
emerged as a natural dynamical length scale in that work, indicating
that the system \lag\ describes some M-theoretic physics.
In \zerobranes\ \lag\ was studied as a low velocity effective
theory appropriate to the heavy \dzb \ of weakly coupled string theory.
Here we  propose \lag\
as the 
most general \imf \ Lagrangian, with at most 
two derivatives which is
invariant under the gauge symmetry and the Supergalilean group
\ref\flume{M. Baake, P. Reinicke, V. Rittenberg, J. Mathm. Phys., 26,
(1985), 1070; R. Flume, Ann. Phys. 164, (1985), 189; M.Claudson,
M.B.Halpern, Nucl. Phys. B250, (1985), 689.
}\foot{The gauge invariance is in fact necessary to
supertranslation invariance.  The supergenerators close on gauge
transformations and only satisfy the supertranslation algebra on the
gauge invariant subspace.}.  It would be consistent with our assumption
that matrix \dzb \ are the only degrees of freedom of \mt \ , to write a
Lagrangian with higher powers of first derivatives.  We do not know if
any such Lagrangians exist which preserve the full symmetry of the \imf \
.  What is at issue here is eleven dimensional Lorentz invariance.  In
typical \imf \ field theories, the naive classical Lagrangian for the
positive longitudinal momentum modes is renormalized by the decoupled
infinite frequency modes.  The criterion which determines the \imf \ 
Lagrangian is invariance under longitudinal boosts and null plane
rotating Lorentz tranformations (the infamous angular conditions).
Apart from simplicity, our main reason for suggesting the 
Lagrangian \lag \ is that we have found some partial evidence that the
large $N$ limit of the quantum theory it defines is indeed Lorentz
invariant.
A possible line of argument systematically leading to \lag\
is discussed in Section 9.

Following \zerobranes\ , let us rewrite the action in units 
in which the 11-D Planck length is $1$. Using \Relevg \ and
\lelevls \ the change of
units is easily made and one finds
\eqn\resclag{L= \tr \biggl[ {1 \over 2R} D_t{Y^i} D_t{Y^i}
- {1\over 4} R\ [Y^i,Y^j]^2 - {\theta^T}  D_t{\theta} - R\ {\theta^T}
\gamma_i [\theta ,Y^i] \biggr]}
where $Y={X\over g^{1/3}}$. We have also changed the units of time to 11-D
Planck units.  We have restored the gauge field ($\partial_t \rightarrow
D_t = \partial_t + i A$) to this expression
(previously we were in $A = 0$ gauge) in order to emphasize that the
SUSY transformation laws (here $\epsilon$ and $\epsilon^{\prime}$ are
two independent $16$ component
anticommuting SUSY parameters),
\eqn\sutrans{\delta X^i = - 2{\epsilon}^T \gamma^i \theta}
\eqn\sutransb{\delta \theta = \ha  \biggl[ D_t X^i \gamma_i +
\gamma_- + \ha [X^i \ ,X^j ]\ \gamma_{ij} \biggr] \epsilon + \epsilon^{\prime}}
\eqn\sutransc{\delta A = - 2{\epsilon^T} \theta}
involve a gauge transformation.  As a result, the SUSY algebra closes on
the gauge generators, and only takes on the form \supalg \ when applied
to gauge invariant states.  

The Hamiltonian has the form
\eqn\ham{
H=R\ \tr \left\{{{{ \Pi_i \Pi_i}\over 2}
+{1\over 4}\ [Y^i,Y^j]^2}+{\theta^T} \gamma_i [\theta,Y^i]
\right\}}
where $\Pi$ is the canonical conjugate to $Y$.  Note that in the limit
$R\rightarrow\infty$, all finite energy states of this Hamiltonian have
infinite energy.  We will be interested only in states whose energy
vanishes like ${1\over N}$ in the large $N$ limit, so that this factor
becomes the inverse power of longitudinal momentum which we expect for
the eigenstates of a longitudinal boost invariant system.  Thus, in the 
correct \imf \ limit, the only relevant
asymptotic states of the Hamiltonian should
be those whose energy is of order ${1\over N}$.  We will exhibit a class
of such states below, the supergraviton scattering states.  The
difficult thing will be to prove that their S-matrix elements depend
only on ratios of longitudinal momenta, so that they are longitudinally
boost invariant.

 To understand how this system represents ordinary
particles we note that when the $Y's$ become large the
commutator term in $H$ becomes very costly in energy. Therefore for large
distances the finite energy configurations lie on the flat directions along
which the commutators vanish. In this system with $16$ 
supercharges, \foot{The sixteen supercharges which anticommute to the longitudinal
momentum act only on the center of mass of the system and play no role
in particle interactions.}\ 
these classical zero energy states are in fact exact supersymmetric
states of the system.  In contrast to field theory, the
continuous parameters which describe these states (the Higgs VEVs in the
language of SYM theory) are not vacuum superselection parameters, but
rather collective coordinates.  We must compute their quantum wave
functions rather than freeze them at classical values.
They are however the slowest modes in
the system, so that we can integrate out the other degrees of freedom to
get an effective SUSY quantum mechanics of these modes alone. 
We will study some aspects of this effective dynamics below.

Along the flat directions the $Y^i$ are
simultaneously diagonalizable. The diagonal matrix elements are the coordinates
of the \dzb. When the $Y$ are small the cost in energy for a noncommuting
configuration is not large. Thus for small distances there is no interpretation
of the configuration space in terms of ordinary positions.  Classical
geometry and distance are only sensible concepts in this system in
regions far out along one of the flat directions.  We will refer to this
as the long distance regime.  In the short distance regime, we have a
noncommutative geometry.
Nevertheless the full
Hamiltonian \ham \ has the usual Galilean symmetry. To see this we define the
center of mass of the system by
\eqn\cofm{Y(c.m.) = {1 \over N} \tr Y}

A transverse translation is defined by adding a multiple of the identity to
$Y$. This has no effect on the commutator term in $L$ because the identity
commutes with all $Y$. Similarly rotational invariance is manifest.

The center of mass momentum is given by
\eqn\Pcm{
P(c.m.) = \tr \Pi = {N \over R}\dot Y(c.m.)}

Using $\p = N/R$ gives the usual connection between transverse velocity and
transverse momentum
\eqn\Pcmm{
 {1\over \p}  P(c.m.)=\dot Y(c.m.)}

A Galilean boost is defined by adding a multiple of the identity to
$\dot Y$. We leave it to the reader to show that this has no effect on the
equations of motion. This establishes the Galilean invariance of $H$. The
Supergalilean invariance is also completely unbroken.
The alert reader may be somewhat unimpressed by some of these
invariances, since they appear to be properties of the center of mass
coordinate, which decouples from the rest of the dynamics.  Their real
significance will appear below when we show that our system possesses
multiparticle asymptotic states, on which these generators act in the
usual way as a sum of single particle operators.

%
%
\newsec{\bf A Conjecture}

Our conjecture is thus that
M-theory formulated in the \imf \ is exactly equivalent to the $N \to \infty$
limit of the supersymmetric quantum mechanics described by the Hamiltonian
\ham. The calculation of any physical quantity in M-theory can be reduced to
a calculation in matrix quantum mechanics followed by an extrapolation to large
$N$. In what follows we will offer evidence
for this surprising conjecture.

Let us begin by examining the single particle spectrum of the theory.
For $N=1$ the states  with
$p_{\perp} =0$ are just those of a single D0 brane at rest.
The states form a representation of the algebra of the 16 $\theta 's$ with
$2^8$ components. These states have exactly the quantum numbers of the 256
states of the supergraviton. For non-zero $p_{\perp}$ the energy of the
object is
\eqn\egy{
E={R \over 2}{p_{\perp}^2}= {{p_{\perp}^2\over 2\p}}}

For states with $N>1$ we must study the $U(N)$ invariant
 Schroedinger equation arising from \ham. $H$ can easily be separated in
terms of center of mass and relative motions. Define
\eqn\yrel{
Y = {{Y(c.m.)}\over N}I + \yr}
where $I$ is the unit matrix and $\yr$ is a traceless matrix in the adjoint of
$SU(N)$ representing relative motion. The Hamiltonian then has the form
\eqn\hrel{
H = H_{c.m.}+H_{rel}}

with
\eqn\hcm{
 H_{c.m.}={P(c.m.)^2 \over 2\p}}

The Hamiltonian for the relative motion is the dimensional reduction of the
supersymmetric 10-D Yang Mills Hamiltonian. Although a direct proof based on
the Schroedinger equation has not yet been given, duality detween IIA strings
and M-theory requires the relative Schroedinger equation to have  normalizable
threshold bound states with zero energy~\w . The bound system must have
exactly the quantum numbers of the 256 states of the supergraviton. For these
states the complete energy is given by \hcm. Furthermore these states are BPS
saturated. No other normalizable bound states can occur. Thus we find that the
spectrum of stable single particle excitations of \ham \ is exactly the
supergraviton spectrum with the correct dispersion relation to describe
massless 11 dimensional particles in the \imf.

Next let us turn to the spectrum of widely separated particles. That a simple
quantum mechanical Hamiltonian like \ham \ should be able to describe
arbitrarily
many well separated particles is not at all evident and would certainly be
impossible without the special properties implied by supersymmetry. Begin by
considering commuting block diagonal matrices of the form
\eqn\split{
\pmatrix{Y^i_1 &0&0&\ldots \cr
         0&Y^i_2&0\ldots \cr
         0&0&Y^i_3 \ldots \cr
         \vdots \cr}}
where $Y^i_a$ are $N_a\times N_a$ matrices and $N_1+N_2 +\ldots+N_n =N$. For
the
moment suppose all other elements of the $Y's$ are constrained to vanish
identically. In this case the Schroedinger equation obviously separates into
$n$
uncoupled Schroedinger equations for the individual block degrees of freedom.
Each equation is identical to the original Schroedinger equation for the system
of $N_a$ \dzb. Thus the spectrum of this truncated system includes
collections of noninteracting supergravitons.

Now let us suppose the supergravitons are very distant from one another. In
other words for each pair the relative distance, defined by
\eqn\sep{
R_{a,b}=|{\tr Y_a \over N_a}-{\tr Y_b \over N_b}|}
is asymptotically large. In this case the commutator terms in \ham \ cause the
off diagonal blocks in the $Y's$ to have very large potential energy
proportional to $R_{a,b}^2$. This effect can also be thought of as the Higgs
effect giving mass to the broken generators (``W-bosons") of $U(N)$ when the
symmetry is broken to $U(N_1) \times U(N_2) \times U(N_3) \ldots$. Thus one
might naively expect the off diagonal modes to leave the spectrum of very
widely separated supergravitons unmodified. However this is not correct in a
generic situation. The off diagonal modes behave like harmonic oscillators with
frequency of order $R_{a,b}$ and their zero point energy will generally give
rise to a potential energy of similar magnitude. This effect would certainly
preclude an interpretation of the matrix model in terms of well separated
independent particles.

Supersymmetry is the ingredient which rescues us. In a well known way
the fermionic partners of the off diagonal bosonic modes exactly cancel the
potential due to the bosons, leaving exactly flat directions.   We know
this from the nonrenormalization theorems for supersymmetric quantum
mechanics with $16$ supergenerators \flume .  The effective Lagrangian
for the collective coordinates along the flat directions must be
supersymmetric and the result of \flume \ guarantees that up to terms
involving at most two derivatives, the Lagrangian for these coordinates
must be the dimensional reduction of $U(1)^n$ SYM theory, where $n$ is
the number of blocks ({\it i.e.} the number of supergravitons).  This is
just the Lagrangian for free motion of these particles.  Furthermore,
since we are doing quantum mechanics, and the analog of the Yang Mills
coupling is the dimensional quantity $l_p^3$, the coefficient of the
quadratic term is uncorrected from its value in the original Lagrangian.

There are residual virtual effects at order $p^4$ from these heavy
states which are the source of parton interactions.  
Note that the off diagonal modes are manifestly nonlocal.  The apparent
locality of low energy physics in this model must emerge from a complex
interplay between SUSY and the fact that the frequencies of the
nonlocal degrees of freedom become large when particles are separated. 
We have only a limited understanding of this crucial issue, but in the
next section we will provide some evidence that local physics is
reproduced in the low energy, long distance limit.

The center of mass of a block of size $N^{(a)}$ is defined by
equation \yrel \ .  It is easy to see that the Hamiltonian for an
asymptotic multiparticle state, when written in terms of center of mass
transverse momenta, is just
\eqn\asyham{H_{asymp} = \sum_a {R {\bf P^{(a)}}^2 \over N^{(a)}} =
\sum_a { {\bf P^{(a)}}^2 \over p_{11}^{(a)}} }
Note that the dispersion relation for the asymptotic particle states has
the fully eleven dimensional Lorentz invariant form.  This is
essentially due to the BPS nature of the asymptotic states.  For large
relative separations, the supersymmetric quantum state corresponding to
the supersymmetric classical flat direction in which the gauge symmetry
is \lq\lq broken\rq\rq into $n$ blocks, will be precisely the product
of the threshold boundstate wave functions of each block subsystem.
Each individual block is a BPS state. Its dispersion relation follows
{}from the SUSY algebra and is relativistically invariant even when ({\it
e.g.} for finite $N$) the full system is not.

  We also note
that
the statistics of multi supergraviton states comes out correctly because
of the residual block permutation gauge symmetry of the matrix model.
When some subset of the blocks are in identical states, the original
gauge symmetry instructs us to mod out by the permutation group, picking
up minus signs depending on whether the states are constructed from an
odd or even number of Grassmann variables.  The spin statistics
connection is the conventional one.

Thus, the large $N$ matrix model contains the Fock space of asymptotic
states of eleven
dimensional supergravity, and the free propagation of particles is
described in a manner consistent with eleven dimensional Lorentz
invariance.  The field theory Fock space is however embedded in a system
which, as we shall see, has no ultraviolet divergences.  Particle
statistics
is embedded in a continuous gauge symmetry.  We find the emergence of
field theory as an approximation to an elegant finite structure one of
the most attractive features of the matrix model approach to \mt .

%
%
\newsec{\bf Long Range Supergraviton Interactions}

The first uncancelled interactions in the matrix model occur in the effective
action at order
${\dot y}^4$ where $\dot y$ is the velocity of the supergravitons
\zerobranes. These
interactions are calculated by thinking of the matrix model as
SYM theory and computing Feynman diagrams. At one loop one finds
an induced quartic term in the velocities which corresponds to an induced
$F_{\mu \nu}^4$ term. The precise term for two \dzb \ is given by
\eqn\precise{
A [\dot y(1)-\dot y(2)]^4\over R^3r^7}
where $r$ is the distance between the \dzb \ and $A$ is a coefficient
of order one which can be extracted from the results of \zerobranes.
This is the longest range term which governs the
interaction between the \dzb \ as $r$ tends to infinity. Thus the effective
Lagrangian governing the low energy long distance behavior of the pair is
\eqn\leff{
L = {{\dot y(1)}^2 \over 2R}+{{\dot y(2)}^2 \over 2R} - A{[\dot y(1)-\dot
y(2)]^4\over R^3r^7}}

The calculation is easily generalized to the case of two well separated groups
of $N_1$ and $N_2$ \dzb \ forming bound states. Keeping only the leading terms
for large $N$ (planar graphs) we find

\eqn\leffn{L = {N_1{\dot y(1)}^2 \over 2R}+{N_2{\dot y(2)}^2 \over 2R}
- A N_1N_2{[\dot y(1)-\dot y(2)]^4\over {R^3 r^7}}}

To understand the significance of \leffn \ it is first useful to translate it
into an effective Hamiltonian. To leading order in inverse powers of $r$ we
find
\eqn\hef{
H_{eff}= { p_{\perp}(1)^2\over 2\p(1)}+{ p_{\perp}(2)^2\over 2\p(2)}
         +A\left[{p_{\perp}(1) \over \p(1)}-{p_{\perp}(2) \over \p(2)}\right]^4
{\p(1)\p(2) \over r^7 R}}

{}From \hef \ we can compute a scattering amplitude in Born approximation.
Strictly speaking the scattering amplitude is defined as a 10-D amplitude
in the compactified theory. However it contains information about the 11-D
amplitude in the special kinematic situation where no longitudinal momentum is
exchanged. The relation between the 10-D amplitude and 11-D amplitude at
vanishing $\Delta \p$ is simple. They are essentially the same thing except for
a factor of $R$ which is needed to relate the 10 and 11 dimensional phase space
volumes. The relation between amplitudes is $A_{11}=RA_{10}$. Thus from
\hef \ we
find the 11-D amplitude
\eqn\amp{
A\left[{p_{\perp}(1) \over \p(1)}-{p_{\perp}(2) \over \p(2)}\right]^4
{\p(1)\p(2) \over r^7 }}

The expression in \amp \ is noteworthy for several reasons.
First of all the factor $r^{-7}$
is the 11-D  green function (in space)
after integration over $\x$. In other words it is
the  scalar Feynman propagator for vanishing longitudinal momentum
transfer. Somehow the simple matrix Hamiltonian knows about massless
propagation in 11-D spacetime. The
remaining momentum dependent factors are exactly what is needed to make
\amp \ identical to the single (super) graviton exchange diagram in 11-D.\foot{
In \zerobranes \ the amplitude was
computed for \dzb \ which
have momenta orthogonal to their polarizations (this was not stated
explicitly there but was implicit in the choice of boundary state).  The spin dependence of the amplitude is
determined by the supersymmetric completion of the $v^4 \over r^7$
amplitude, which we have not computed.  In principle, this gives another
check of eleven dimensional Lorentz invariance.  We suspect that the
full answer follows by applying the explicit supersymmetries of the
light cone gauge to the amplitude we have computed. }
Even
the coefficient is correct. This is closely related
to a result reported in \zerobranes\ where it was shown that the annulus
diagram
governing the scattering of two \dzb \ has exactly the same form at very small
and very large distances, which can be understood by 
noting that only BPS states contribute to this process on the annulus.
This, plus the usual relations between
couplings and scales in Type IIA string theory and \mt\ guarantee that
we obtain the correct normalization of eleven dimensional graviton
scattering in SUGRA.
In the weak coupling limit, very long distance behavior is governed by single
supergraviton exchange while the ultra short distances are governed by the
matrix model. In \zerobranes\ the exact equivalence between the leading
interactions computed in these very different manners was recognized
but
its meaning was not clear. Now we see that it is an important consistency
criterion in order for the matrix model to describe the \iml \ of M-theory.

Let us next consider possible corrections to the effective action coming from
higher loops. In particular, higher loops can potentially correct the quartic
term in velocities. Since our interest lies in the large $N$ limit we may
consider the leading (planar) corrections. Doing ordinary large $N$ counting
one finds that the ${\dot y}^4$ term may be corrected by a factor which is a
function of the ratio $N/r^3$. Such a renormalization by $f(N/r^3)$ could be
dangerous. We can consider several  cases which differ in the behavior of $f$
as
${N\over r^3}$ tends to infinity. In the first two the function tends to zero
or infinity. The meaning of this would be that the coupling to gravity is
driven either to zero or infinity in the \iml. Either behavior is intolerable.
 Another possibility is that the function $f$  tends to a constant not
equal to
$1$. In this case the
gravitational coupling constant is renormalized by a constant factor. This is
not supposed to occur in M-theory. Indeed supersymmetry is believed to protect
the gravitational coupling from any corrections. The only other possibility is
that $f\to 1$. The simplest way in which this can happen is if there are no
corrections at all other than the one loop term which we have discussed.

We believe that there is a nonrenormalization theorem for this term
which can be proven in the context of SUSY quantum mechanics with $16$
generators.  The closest thing we have been able to find in the
literature is a nonrenormalization theorem for the ${F_{\mu
\nu}}^4$ term in the action of ten dimensional string theory\foot{We
thank C. Bachas for pointing out this theorem to us.} which has been
proven by
Tseytlin \ref\tseytlin{A.A. Tseytlin,``On $SO(32)$ Heterotic - type I
Superststring Duality in Ten Dimensions" Imperial /TP/95-96/6~
hep-th/9510173\semi ``Heterotic - Type I Superstring Duality and Low -Energy
Effective Actions,'' Imperial /TP/95-96/16~  hep-th/9512081.}.
the quantum mechanical context, we believe that it is true and that the 
scattering of two supergravitons at large transverse 
distance and zero longitudinal momentum is exactly given in the matrix
model by low energy 11-D
supergravity perturbation theory.  M. Dine\ref\dine{M. Dine, work in
progress.} has constructed the outlines
of an argument which demonstrates the validity of the nonrenormalization 
theorem.

We have considered amplitudes in which vanishing longitudinal momentum is
exchanged. Amplitudes with nonvanishing exchange of $\p$ are more complicated.
They correspond to processes in 10-D in which RR charge is exchanged. Such
collisions involve rearrangements of the \dzb \ in which the collision
transfers \dzb \ from one group to the other. We are studying
such processes but we have no definitive results as yet.

We have thus presented some evidence that the dynamics of the matrix
model respects eleven dimensional Lorentz invariance.  If this is
correct then the model reduces exactly to supergravity at low energies.
It is clear however that it is much better behaved in the ultraviolet
than a field theory. 
At short distances, as shown extensively in \zerobranes \
restoration of the full matrix character of the variables cuts off all
ultraviolet divergences.  The correspondence limit by which \mt  \  reduces
to supergravity indicates that we are on the right track.

\newsec{The Size of a Supergraviton}

As we have pointed out in section 2 the holographic principle requires the
transverse size of a system to grow with the number of constituent partons. It
is therefore of interest to estimate the size of the threshold bound state
describing a supergraviton of longitudinal momentum $N/R$. According to the
holographic principle the radius should grow like $N^{1/9}$ in 11-D
Planck units. We will use a mean field approximation in which we study the wave
function of one parton in the field of $N$ others. We therefore consider the
effective Lagrangian \leffn \ for the case $N_1 =1, N_2=N$. The action
simplifies for $N>>1$ since in this case the $N$ particle system is much
heavier than the single particle. Therefore we may set its velocity to zero.
The Lagrangian becomes

\eqn\mnfld{
L_1 = {{\dot y}^2 \over 2R}
- N{{\dot y}^4\over {R^3 y^7}}}
where $y$ refers to the  relative coordinate between the two systems.
We can remove all $N$ and $R$  dependence from the action $S=\int L_1
dt$ by scaling

\eqn\rescale{
 \eqalign{y & \rightarrow N^{1/9}y \cr
  t & \rightarrow {N^{2/9}\over R}t ~.}}
The characteristic length, time and velocity ($v= \dot y$) scales are 

\eqn\charscales{
\eqalign{y & \sim N^{1/9} \cr
         t & \sim {N^{2/9}\over R} \cr
         v & \sim {y \over t} \sim N^{-1/9} R ~. } }

That the size of the bound state wave function scales like $N^{1/9}$ is an
indication of the incompressibility of the system when it achieves a
density of order one degree of freedom per Planck area.  This is in
accord with the holographic principle.

This mean field picture of the bound state, or any other description 
of it as a simple cluster, makes the problem
of longitudinal boost invariance mentioned earlier very concrete.
Suppose we consider the scattering of two bound states with $N_1$
and $N_2$ constituents respectively, $N_1 \sim N_2 \sim N$.
The mean field picture
strongly suggests that scattering  will show a characteristic feature
at an impact parameter corresponding to the bound state size $ \sim 
N^{1/9}$.
But this is not consistent with longitudinal boosts which take
$N_1 \rightarrow \alpha N_1, ~N_2 \rightarrow \alpha N_2$. Boost
invariance requires physics to depend only on the ratio $N_1/N_2$, 
or said another way, only on the ratio of the bound state sizes.  
This strongly suggests that a kind of scale invariance 
must be present in the dynamics that is clearly absent in the simple
picture discussed above.
In the string case the scale invariant
world sheet dynamics is crucial for longitudinal boost invariance.

The possibility that partons might form subclusters within the bound
state was ignored in  mean field discussion.  A preliminary discussion
of a hierarchical clustering model with many length scales is presented
in the Appendix.  Note also that wavefunctions of threshold bound states
are power law behaved.

Understanding the dynamics of these bound states well enough to 
check longitudinal boost invariance reliably is an important 
subject for future research. 
%
%
\newsec{Membranes}

In order to be the strong coupling theory of IIA string theory, M-theory must
have membranes in its spectrum. Although in the decompactified limit there are
no truly stable finite energy membranes, very long lived large  classical
membranes  must exist. In this section we will show
how these membranes are
described in the matrix model,
a result first found in \hoppe \foot{We are grateful to
M. Green for pointing out this paper to us when a preliminary version
of this work was presented at 
the Santa Barbara Strings 96 conference.} .
Townsend\townsrem \ first pointed out the  connection between the matrix
description of D0 brane dynamics and the matrix description of membranes,
and speculated that a membrane might be regarded as a collective
excitation of \dzb .  Our conjecture supplies a precise realization of
Townsend's idea.
\nref\zachos{
D. Fairlie, P. Fletcher, \& C. Zachos, 
J. Math. Phys. {\bf 31} (1990) 1088-1094.}
\nref\hoptwo{J. Hoppe, Int. J.
Mod. Phys. A4 (1989) 5235.}

The formulation we will use to describe this connection is a 
version of the methods introduced in \refs{\hoppe,\zachos,\hoptwo}  .

Begin with a pair of unitary operators $U,V$ satisfying the relations
\eqn\algebra{
\eqalign{UV&=e^{2 \pi i \over N}VU \cr
         U^N&=1 \cr
         V^N&=1  ~.}}
These operators can be represented on an $N$ dimensional Hilbert space as clock
and shift operators. They form a basis for all operators in the space. Any
matrix $Z$ can be written in the form
\eqn\expand{
Z=\sum _{n,m=1}^N{Z_{nm}}U^m V^m}
$U$ and $V$ may be thought of as exponentials of canonical variables $p$ and
$q$ :
\eqn\pandq{
\eqalign {U&=e^{ip}\cr V&=e^{iq} ~.}}
where $p,q$ satisfy the commutation relations
\eqn\comm{
[q,p] = {2\pi i \over N}~.}

{}From \expand \ we see that only periodic functions of $p$ and $q$ are
allowed.  
Thus the space defined by these variables is a torus. In fact there is an
illuminating interpretation of these coordinates in terms of the quantum
mechanics of particles on a torus in a strong background magnetic field. The
coordinates of the particle are $p,q$. If the field is strong enough the
existence of a large gap makes it
useful to truncate the space of states to the finite dimensional subspace of
lowest Landau levels. On this subspace the commutation relations \comm \ are
satisfied. The lowest Landau wave packets form minimum uncertainty packets
which occupy an area $\sim 1/N$ on the torus. These wave packets are analogous
to the ``Planckian cells'' which make up quantum phase space. The $p,q$ space is
sometimes called the  noncommuting torus, the quantum torus or the fuzzy torus
.  In fact, for large enough $N$ we can choose other bases of $N$
dimensional Hilbert space which correspond to the lowest Landau levels
of a charged particle propagating on an arbitrary Riemann surface
wrapped by a constant magnetic field. For example in \hoppe , de Wit {\it
et al.} construct the finite dimensional Hilbert space of lowest Landau
levels on a sphere.  This connection between finite matrix models and
two dimensional surfaces is the basis for the fact  that the large
$N$ matrix model contains membranes.  For finite $N$, the model consists
of maps of quantum Riemann surfaces into a noncommuting transverse
superspace, {\it i.e.} it is a model of a noncommuting membrane embedded
in a noncommutative space \foot{Note that it is clear in this context
that membrane topology is not conserved by the dynamics.  Indeed for
fixed $N$ a given matrix can be thought of as a configuration of many
different membranes of different topology.  It is only in the large $N$
limit that stable topological structure may emerge in some situations.}.

In the limit of large $N$ the quantum torus behaves more and more like
classical phase space. The following correspondences connect the two:

\item{1.} The quantum operators $Z$ defined in \expand \ are replaced by their
classical counterparts. Eq. \expand \ becomes the classical Fourier
decomposition of a function on phase space.

\item{2.} The operation of taking the trace of an operator goes over to $N$
times the integral over the torus.
\eqn\trace{
\tr Z \to N\int Z(p,q) dp dq}

\item{3.} The operation of commuting two operators is replaced by
$1/N$ times the classical Poisson bracket
\eqn\Poisson{
[Z,W] \to {1 \over N}[\pa_q {Z}\pa_p {W}-\pa_q {W}\pa_p {Z}]}

We may now use the above correspondence to formally rewrite the matrix model
Lagrangian. We begin by representing the matrices $Y^i$ and $\theta$ as
operator
functions
$Y^i(p,q), \theta (p,q)$. Now apply the correspondences to the two terms in
\resclag . This gives
\eqn\lagnian{
\eqalign{L=&{\p\over2 }\int dpdq \dot{Y^i(p,q)}^2 -
{1\over \p}\int dpdq \ [\pa_q {Y^i}\pa_p {Y^j}-\pa_q
{Y^j}\pa_p {Y^i}]^2 \cr &+fermionic\ terms \cr}}

and a Hamiltonian
\eqn\hamian{
\eqalign{H=&{1\over2\p }\int dpdq \ \Pi_i(p,q)^2 +
{1\over \p}\int dpdq[\pa_q {Y^i}\pa_p {Y^j}-\pa_q
{Y^j}\pa_p {Y^i}]^2 \cr  &+ fermionic\ terms\cr}}

Equation \hamian \ is exactly the standard Hamiltonian for the 11-D supermembrane in
the light cone frame. The construction shows us how to build configurations in
the matrix model which represent large classical membranes. To do so we start
with a classical embedding of a toroidal membrane described by periodic
functions $Y^i(p,q)$. The Fourier expansion of these functions provides us
with a set of coefficients $Y^i_{mn}$. Using \expand \ we then replace the
classical $Y's$ by operator functions of $U,V$. The resulting matrices
represent the large classical membranes.

If the matrix model membranes described above are to correspond to
M-theory membranes their tensions must agree.  Testing this
involves keeping track of the numerical factors of order one
in the above discussion.  We present this calculation in Appendix B
where we show that the matrix model membrane tension exactly agrees with 
the M-theory membrane tension.  This has also been verified by 
Berkooz and Douglas \ref\berkdoug{M. Berkooz and M. R. Douglas, 
`` Five-branes in M(atrix) Theory,'' hep-th/9610236.}
using a different technique.

We do not expect static finite energy membranes  to exist in the uncompactified
limit. Nevertheless let us consider the conditions for such a static solution.
The matrix model equations of motion for static configurations is
\eqn\static{[Y^i,[Y^j,Y^i]]=0 .}

It is interesting to consider a particular limiting case of an infinite
membrane stretched out in the $8,9$ plane for which a formal solution of
\static \ can be found. We first rescale $p$ and $q$
\eqn\rescale{
\eqalign{P&=\sqrt{N}p \cr Q&=\sqrt{N}q \cr [Q,P]&=2 \pi i \cr}}

In the $N \to \infty $ limit the $P,Q$ space becomes an infinite plane. Now
consider the configuration
\eqn\stretch{
\eqalign{Y^8&= R_8 P \cr Y^9&= R_9 Q \cr }}
with all other $Y^i=0$.   $R_i$ is the length of the corresponding
direction, which should of course be taken to infinity.
Since $[Y^8,Y^9]$ is a c-number, eq \static \ is
satisfied. Thus we find the necessary macroscopic membranes require by
M-theory.  This stretched membrane has the requisite ``wrapping
number"  on the infinite plane.  On a general manifold one might
expect the matrix model version of the wrapping number of a membrane on
a two cycle to be
\eqn\wrap{W = {1\over N} Tr\ \omega_{ij}(X^i (p,q)) [X^i , X^j]}
where $\omega$ is the two form associated to the cycle.  This expression
approaches the classical winding number as we take the limit in which
Poisson brackets replace commutators.

Another indication that we have found the right representation of the
membrane comes from studying the supersymmetry transformation properties
of our configuration\foot{This result was derived in collaboration with
N. Seiberg, along with a number of other observations about
supersymmetry in the matrix model, which will appear in a future 
publication.}.  The supermembrane should preserve half of the
supersymmetries of the model.  The SUSY transformation of the fermionic
coordinates is
\eqn\deltatheta{\delta\theta = (P^i \gamma_i + [X^i ,
X^j]\gamma_{ij})\epsilon + \epsilon^{\prime}}
For our static membrane configuration, $P^i = 0$, and the commutator is
proportional to the unit matrix, so we can choose $\epsilon^{\prime}$ to
make this variation vanish.  The unbroken supergenerators are  linear
combinations of the IMF $q_A$ and $Q_{\alpha}$.

It is interesting to contempate a kind of duality and complementarity between
membranes and \dzb. According to the standard light cone quantization of
membranes, the longitudinal momentum $\p $ is uniformly distributed over the
area of the $p,q$ parameter space. This is analogous to the uniform
distribution of $\p$ along the $\sigma$ axis is string theory. As we have seen
the $p,q$ space is a noncommuting space with a basic indivisible quantum of
area. The longitudinal momentum of such a unit cell is $1/N$ of the total. In
other words the unit phase space cells that result from the noncommutative
structure of $p,q$ space are the \dzb\ with which we began.  The \dzb \ and
membranes are dual to one another. Each can be found in the theory of the
other.

The two kinds of branes also have a kind of complementarity. As we have seen,
the configurations of the matrix model which have classical interpretations in
terms of \dzb \ are those for which the $Y's$ commute. On the other hand the
configurations of a membrane which have a classical interpretation are the
extended membranes of large classical area. The area element is the Poisson
bracket which in the matrix model is the commutator. Thus the very classical
membranes are highly nonclassical configurations of \dzb.

In the paper of de Wit, Luscher, and Nicolai \luscher \
 a pathology of membrane theory was reported. It was found
that the spectrum of the membrane Hamiltonian is continuous. The reason for
this is the existence of the unlifted flat directions along which the
commutators vanish. Previously it had been hoped that membranes would behave
like strings and have discrete level structure and perhaps be the basis for a
perturbation theory which would generalize string perturbation theory. In the
present context this apparent pathology is exactly what we want. M-theory
has no
small coupling analogous to the string splitting amplitude. The bifurcation of
membranes when the geometry degenerates is expected to be an order one
process. The matrix model, if it is to describe all of M-theory must
inextricably contain this process. In fact we have seen how
important it is that supersymmetry maintains the flat directions.
A model of a single noncommutative membrane actually contains an entire
Fock space of particles in flat eleven dimensional space time.

Another pathology of conventional membrane theories which we expect to
be avoided in \mt\ is the nonrenormalizability of the membrane world
volume field theory.  For finite $N$, it is clear that ultraviolet
divergences on the world volume are absent because the noncommutative
nature of the space defines a smallest volume cell, just like a Planck
cell in quantum mechanical phase space (but we should emphasize here
that this is a classical rather than quantum mechanical effect in the
matrix model).  The formal continuum limit which gives the membrane
Hamiltonian is clearly valid for describing the classical motion of a
certain set of metastable semiclassical states of the matrix model.
It should not be expected to capture the quantum mechanics of the full
large $N$ limit.  In particular it is clear that the asymptotic
supergraviton states would look extremely singular and have no real
meaning in a continuum membrane formalism.  We are not claiming here to
have a proof that the large $N$ limit of the matrix quantum mechanics
exists, but only that the issues involved in the existence of this limit are
not connected to the renormalizability of the world volume field theory
of the membrane.

There is one last point worth making about membranes. It involves evidence for
11-D Lorentz invariance of the matrix model. We have considered in some detail
the Galilean invariance of the \imf \ and found that it is satisfied. But there
is more to the Lorentz group. In particular there are generators $J^i$ which in
the light cone formalism rotate the light like surface of initial conditions.
The conditions for invarance under these transformations are the notorious
angular conditions. We must also impose longitudinal boost invariance. 
The angular conditions are what makes Lorentz invariance so
subtle in light cone string theory. It is clearly important to determine
if the matrix model satisfies the angular conditions in the large $N$ limit. In
the full quantum theory the answer is not yet clear but at the level of the
classical equations of motion the answer is yes. The relevant calculations
were done by de Wit, Marquard and Nicolai \ref\Lorentz{B.de Wit, V. Marquard
and H. Nicolai ``Area Preserving Diffeomorphisms and Supermembrane Lorentz
Invariance" Commun. Math. Phys. 128, 39 (1990).}. The analysis is too
complicated to repeat here but we can describe the main points.

The equations for classical membranes can be given in covariant form in terms
of
a Nambu-Goto type action. In the covariant form the generators ot the full
Lorentz group are straightforward to write down. In passing to the light cone
frame the expressions for the nontrivial generators become more complicated but
they are quite definite. In fact they can be expressed in terms of the $Y(p,q)$
and their canonical conjugates $\Pi (p,q)$. Finally, using the correspondence
between functions of $p,q$ and matrices we are led to matrix expressions for
the generators. The expressions of course have factor ordering ambiguities but
these, at least formally, vanish as $N \to \infty$. In fact according to
\Lorentz\ the violation of the angular conditions goes to zero as $1/N^2$.
Needless
to say a quantum version of this result would be very strong evidence for our
conjecture.

We cannot refrain from pointing out that the quantum version of the
arguments of \Lorentz \ is apt to be highly nontrivial.  In particular,
the classical argument works for every dimension in which the classical
supermembrane exists, while, by analogy with perturbative string theory, we only
expect the quantum Lorentz group to be recovered in eleven dimensions. 
Further, the longitudinal boost operator of \Lorentz\ is rather trivial
and operates only on a set of zero mode coordinates, which we have not
included in our matrix model.  Instead, we expect the longitudinal boost
generator to involve rescaling $N$ in the large $N$ limit, and thus to
relate the Hilbert spaces of different SUSY quantum mechanics models.
We have already remarked in the previous section 
that, as anticipated in \holo , longitudinal boost invariance is the key
problem in our model.  We expect it to be related to a generalization of
the conformal invariance of perturbative string theory.

\newsec{\bf  Towards Derivation and
Compactification}

In this section we would like to present a 
line of argument which may lead to a proof of the
conjectured equivalence between the matrix model and \mt .
It relies on a stringy extension of the 
conjectured nonrenormalization theorem combined with the possibility
that all velocities in the large $N$ cluster go to zero as $N
\rightarrow \infty$.

Imagine that $R$ stays fixed as $N \rightarrow \infty$. 
Optimistically one might imagine that finite
$R$ errors are as small as in perturbative 11D supergravity, meaning that they
are suppressed by powers of $k_{11} R$ or even $\exp(-k_{11} R)$ where
$ k_{11}$
is the center of mass longitudinal momentum transfer.  
So for $k_{11} \sim 1/l_p$ we could imagine $R$ fixed at a macroscopic scale
and have very tiny errors.  The mean field estimates discussed in
section 7 give the velocity $v \sim N^{-1/9} R$, which, with $R$ fixed,
can be made
arbitrarily small at large $N$. Although it is likely that the structure
of the large $N$ cluster is more complicated than the mean field 
description, it is possible that this general property of vanishing velocities
at large $N$ continues to hold.  In particular, in Appendix A we present
arguments that the velocities of the coordinates along some of the classical
flat directions of the potential are small.  We suspect that this can be
generalized to all of the flat directions.  If that is the case then the
only high velocities, would be those associated with the \lq\lq core
\rq\rq wave function of Appendix A.  The current argument assumes that
the amplitude for the core piece of the wave function vanishes in the
large $N$ limit.

Nonabelian field strength is 
the correct generalization of velocity 
for membrane type field configurations like those
discussed in Section 8. For classical configurations 
at least these field strengths
are order $1/N$ and so are also small.

We have previously conjectured that the $v^4$ terms in the quantum 
mechanical effective
action are not renormalized beyond one loop. For computing the eleven
dimensional supergravity amplitude, we needed this result in the matrix
quantum mechanics, but is possible that this
result holds in the full string perturbation 
theory.  For example, the excited open
string states can be represented as additional non-BPS fields in the
quantum mechanics.  These do not contribute to the one loop $v^4$ term
because they are not BPS.  Perhaps they do not contribute to higher
loops for related reasons.

If these two properties hold then the conjecture follows.
The scattering of large $N$ clusters
of D-0-branes can clearly be computed at small $g$ (small $R$) using 
the quantum mechanics.  But these
processes, by assumption, only involve low velocities
independent of $g$,
and so only depend on the $v^4$ terms in the 
effective action which, by the stringy extension of the
nonrenormalization
theorem, would not receive $g$ corrections.  So the
same quantum mechanical answers would be valid at large $g$ (large $R$).

This would prove the conjecture.

{}From this point of view we have identified a subset of 
string theory processes (large $N$ D-0-brane scattering) which are unchanged
by stringy loop corrections and so are computable at strong coupling.

If this line of argument is correct, it gives us an unambiguous 
prescription for
compactification.  We take the quantum mechanics which describes 0-brane
motion at weak string coupling 
in the compactified space and then follow it to strong coupling.
This approach to compactification requires us to add extra degrees of
freedom in the compactified theory.  We will discuss an alternative
approach in the next section.

For toroidal compactifications, it is clear at weak coupling that
one needs to keep the strings which wrap any number of times 
around each circle.  These unexcited
wrapped string states are BPS so they do contribute to the $v^4$
term and hence must 
be kept.
In fact these are the states which, in the annulus diagram, 
correct the power law
in graviton scattering to
its lower dimensional value, and are crucial in implementing the various
T dualities.

To be specific let us discuss the case of one coordinate $X^9$
compactified on a circle on radius $R_9$.  Here we should keep the 
extra string winding states around $X^9$. An efficient way to keep
track of them is to T dualize the $X^9$ circle.  This converts
D-0-branes to D-1-branes and winding modes to momentum modes.
The collection of $N$ D-1-branes is described
by a 1+1 dimensional
$SU(N)$ Super Yang Mills quantum field theory with coupling
$g^2_{\rm SYM}=R^2/(R_9 l_p^3) $
on a space  of T dual radius $R_{\rm SYM}=l_p^3/(R R_9)$. The
dimensionless effective coupling of the Super Yang Mills theory
is then $g^2_{\rm SYM} R_{\rm SYM}^2 = (l_p/R_9)^3$ which is 
independent of $R$.
For p-dimensional tori we get systems of D-p-branes described
by p+1 dimensional $SU(N)$ Super Yang Mills theory.
Related issues
have also recently been discussed in \ref\taylor{W. Taylor,
``D-brane field theory on compact spaces,'' hep-th/9611042}.

For more general compactifications the rule would be
to keep every BPS state which contributes to the $v^4$ term at large $N$. 
We are currently investigating such compactifications, including ones
with less supersymmetry.

The line of argument presented in this section raises a number of questions.
Is it permissible to hold $R$ finite, or to let it grow very slowly with $N$?
Are there nonperturbative corrections to the $v^4$ term?  Large
$N$ probably prohibits instanton corrections in the quantum mechanics,
but perhaps not in the full string theory.  This might be related to
the effect of various wrapped branes in compactified theories.

Does the velocity stay low?  A key problem here is that in the mean
field  theory 
cloud the 0-branes are moving very slowly.  If  two 0-branes encounter
each other, their relative velocity is much less than the typical velocity in
a bound pair ($v \sim R$).  It seems that the capture cross section 
to go into the pair bound state should be very large.  Why isn't there
clumping into pairs?  One factor which might come into play is the following.
If the velocity is very low the de Broglie wavelength of 
the particles might be comparable to the whole cluster
(this is true in mean field), so their could be
delicate phase correlations across the whole cluster--some kind of 
macroscopic quantum coherence. Whenever a pair is trying to form, 
another 0-brane might get between them and disrupt them.  This extra
coherent complexity might help explain the Lorentz invariance puzzle.

\newsec{\bf Another Approach to Compactification}

The conjecture which we have presented refers to an exact formulation of
\mt \ in uncompactified eleven dimensional spacetime.  It is tempting to
imagine that we can regain the compactified versions of the theory as
particular collections of states in the large $N$ limit of the matrix
model.  There is ample ground for suspicion that this may not be the
case, and that degrees of freedom that we have thrown away in the
uncompactified theory may be required for compactification.  Indeed, in
IMF field theory the only general method for discussing theories with
moduli spaces of vacua is implementable only when the vacua are visible
in the classical approximation.  Then, we can shift the fields and do
IMF quantization of the shifted theory.  Different vacua correspond to
different IMF Hamiltonians for the same degrees of freedom.  The
proposal of the previous section is somewhat in the spirit of IMF field
theory. Different Hamiltonians, and indeed, different sets of degrees of
freedom, are required to describe each compactified vacuum.

We have begun a preliminary investigation of the alternative hypothesis,
that different compactifications are already present in the model we
have defined.  This means that there must be collections of states
which, in the large $N$ limit, have S matrices which completely decouple
{}from each other.  Note that the large $N$ limit is crucial to the
possible existence of such superselection sectors.  The finite $N$
quantum mechanics cannot possibly have superselection rules.
Thus, the only way in which we could describe compactifications for finite
$N$ would be to add degrees of freedom or change the Hamiltonian.
We caution the reader that the approach we will describe below is 
very preliminary and highly
conjectural.

This approach to compactification is based on the idea that there is a
sense in which our system defines a single \lq\lq noncommuting
membrane\rq\rq .  Consider compactification of a membrane on a circle.
Then there are membrane configurations in which the embedding
coordinates do not transform as scalars under large diffeomorphisms of
the membrane volume but rather are shifted by large diffeomorphisms of
the target space.  These are winding states.  A possible approach to
identifying the subset of states appropriate to a particular
compactification, is to first find the winding states, and then find all
states which have nontrivial scattering from them in the large $N$
limit.  In fact, our limited study below seems to indicate that all
relevant states, including compactified supergravitons, can be thought
of as matrix model analogs of membrane winding states.

Let us consider compactification of the ninth transverse direction on a
circle of radius $2\pi R_9$.
A winding membrane is a configuration which satisfies
\eqn\wind{X^9 (q,p + 2\pi) = X^9 (q,p) + 2\pi R_9}
and the winding sector is defined by a path integral over configurations
satisfying this boundary condition.
A matrix analog of this is
\eqn\matwind{e^{- i Nq} X^9 e^{iNq}= X^9 + 2\pi R_9}
It is easy to see (by taking the trace) that this condition cannot be
satisfied for finite $N$.  However, if we take the large $N$ limit in
such a way that 
$q \rightarrow {\sigma \over N} \otimes {\bf 1}_{M\ {\rm X}\ M}$, 
with $\sigma$ an angle variable, then this equation can be satisfied,
with
\eqn\matwindb{X^9 = {2\pi R_9 \over i}{\partial\over \partial\sigma}
\otimes {\bf 1}_{M\ {\rm X}\ M} + x^9 (\sigma )}
where $x^9$ is an $M {\rm X} M$ matrix valued function of the angle
variable.  The other transverse bosonic coordinates, and all of the
$\theta$s are  $M {\rm X} M$ matrix valued functions of $\sigma$.
 These equations should be thought of as limits of finite
matrices.  Thus $2\pi R_9 {\cal P} \equiv 
{2\pi R_9 \over i}{\partial\over \partial\sigma}$ can
be thought of as the limit of the finite matrices ${\rm diag (-2\pi P
R_9 \ldots 2\pi P R_9)}$, with $\sigma$ the obvious tridiagonal matrix in
this representation.  The total longitudinal momentum of such a
configuration is ${(2P + 1)M \over R}$, and the ratio $M/P$ is an
effectively continuous parameter characterizing the states in the large
$N$ limit.  We are not sure of the meaning of this parameter.

To get a feeling for the physical meaning of this proposal, we examine
the extreme limits of large and small $R_9$.  
For large $R_9$ it is convenient to work in the basis where ${\cal P}$
is diagonal.  If we take all of the coordinates $X^i$ independent of
$\sigma$, then our winding membrane approaches a periodic array of $(2P + 1)$
collections of \dzb \ , each with longitudinal momentum $M/R$.  We can find a
solution of the BPS condition by putting each collection into the $M$ zero
brane threshold bound state wave function.  For large $R_9$
configurations of the $X^i$ which depend on $\sigma$ have very high
frequency, and can be integrated out.  Thus, in this limit, the BPS
state in this winding sector is approximately a periodic array of
supergravitons.   We identify this with the compactified supergraviton state.
This state will have the right long range gravitational
interactions (at scales larger than $R_9$) 
in the eight uncompactified dimensions.  To obtain the correct
decompactified limit it would appear that we must rescale $R$, the
radius of the longitudinal direction by $R \rightarrow (2P + 1) R$ as we
take $P$ and $M$ to infinity.  With this rescaling, all trace of the
parameter $M/P$ seems to disappear in the decompactification limit.

For small $R_9$, our analysis is much less complete.  However, string
duality suggests an approximation to the system in which we keep only
configurations with $M = 1$ and $P \rightarrow \infty$.  
In this case the $\sigma$ dependence of $X^9$ is pure gauge, and the
$X^i$ all commute with each other.  
The matrix model Hamiltonian becomes the Hamiltonian of the
Green-Schwarz type IIA string:
\eqn\stringham{H \rightarrow \int d\sigma (\dot{{\bf X}})^2 + ({\partial
{\bf X} \over \partial\sigma})^2 + \theta^T \gamma_9 {\partial\theta
\over \partial\sigma}}

As in previous sections, we will
construct multi wound membrane states by making large block
diagonal matrices, each block of which is the previous single particle
construction.  Lest such structures appear overly baroque, we remind
the reader that we are trying to make explicit constructions of the
wave functions of a strongly interacting system with an infinite number
of degrees of freedom.   For large $R_9$ it is fairly clear that
the correct asymptotic properties of
multiparticle states will be guaranteed by the BPS condition (assuming
that everything works as conjectured in the uncompactified theory).

If our ansatz is correct for small $R_9$, 
it should be possible to justify the neglect
of fluctuations of the matrix variables away from the special forms we
have taken into account, as well as to show that the correct
string interactions (for multistring configurations defined by the sort
of block diagonal construction we have used above) are obtained from the
matrix model interactions.  In this connection it is useful to note that
in taking the limit from finite matrices, there is no meaning to the
separation of configurations into winding sectors which we have defined
in the formal large $N$ limit.  In particular, $X^9$ should be allowed
to fluctuate.  But we have seen 
that shifts of $X^9$ by functions of $\sigma$ are pure gauge, so that 
all fluctuations around the configurations which we have kept,
give rise to higher derivative world sheet interactions.  Since the
$P\rightarrow \infty$ limit is the world sheet continuum limit we should
be able to argue that these terms are irrelevant operators in that
limit.   We have less understanding about how the sum over world sheet
topologies comes out of our formalism, but it is tempting to think that
it is in some way connected with the usual topological expansion of
large $N$ matrix models.  In the appendix we show that in eleven
dimensions, dimensional analysis guarantees the dominance of planar
graphs in certain calculations.  Perhaps, in ten dimensions, the small
dimensionless parameter ${{R_9}\over l_p}$ must be scaled with a
power of $N$ in order to obtain the limit of the matrix model which
gives IIA string theory.  

These ideas can be extended to compactification on multidimensional
tori.  A wrapping configuration of a toroidal membrane can be
characterized by describing the cycles on the target torus
on which the $a$ and $b$ cycles of the membrane are mapped.  This
parametrization is redundant because of the $SL(2,Z)$ modular invariance
which exchanges the two membrane cycles.  We propose that the analogs of
these wrapping states, for a $d$-torus defined by modding out $R^d$  by
the shifts $X^i \rightarrow X^i + 2\pi R^i_a$, is defined by the
conditions
\eqn\dwrap{e^{-i\sigma} X^i_{({\bf m,n})} e^{i\sigma} =  X^i_{({\bf
m,n})} + 2\pi R^i_a n^a}
\eqn\dwrapb{e^{2\pi i{\cal P}} X^i_{({\bf m,n})} e^{- 2\pi i {\cal P} } =  X^i_{({\bf
m,n})} + 2\pi R^i_a m^a}
where ${\bf n}$ and ${\bf m}$ are d-vectors of integers.
The solutions to these conditions are
\eqn\dwrapsol{ X^i_{({\bf m,n})} = \bigl[ 2\pi R^i_a n^a {\cal P} +
R^i_a m^a \sigma \bigr] \otimes 1_{M {\rm X} M} +  x^i_{({\bf m,n})}
(\sigma )}
where the $x^i$ are periodic $M {\rm X} M$ matrix valued functions of
$\sigma$.   The fermionic, and noncompact, coordinates are also matrix
valued functions of $\sigma$.  
 
In order to discuss more complicated compactifications, we would have to
introduce coordinates and find a group of large diffeomorphisms
associated with one and two cycles around which membranes can wrap.
Then we would search for embeddings of this group into the large $N$
gauge group.  Presumably, different coordinate systems would correspond
to unitarily equivalent embeddings.  We can even begin to get a glimpse
of how ordinary Riemannian geometry would emerge from the matrix system.
If we take a large manifold which breaks sufficient supersymmetries, the
effective action for supergravitons propagating on such a manifold would
be obtained, as before, by integrating over the off diagonal matrices.
Now however, the nonrenormalization theorem would fail, and the kinetic
term for the gravitons would contain a metric.  The obvious conjecture
is that this is the usual Riemannian metric on the manifold in question.
If this is the case, our prescription for compactification in the
noncommutative geometry of the matrix model, would reduce to ordinary
geometry in the large radius limit.

A question which arises is whether the information about one and two
cycles is sufficient to characterize different compactifications.  We
suspect that the answer to this is no.  The moduli of the spaces that
arise in string theoretic compactifications are all associated with the
homology of the space, but in general higher dimensional cycles ({\it
e.g.} three cycles in Calabi-Yau three folds) are necessary to a
complete description of the moduli space.  Perhaps in order to capture
this information we will have to find the correct descriptions of five
branes in the matrix model.   If the theory really contains low energy
SUGRA, then it will contain solitonic five branes, but it seems to us
that the correct prescription is to define five branes as the D-branes
of membrane theory.  We do not yet understand how to introduce this
concept in the matrix model. 

Finally, we would like to comment on the relation between the
compactification schemes of this and the previous sections.  For a
single circle, if we take
$P$ to infinity, and substitute the formula \matwindb \ into the matrix
model Hamiltonian (as well as the prescription that all other
coordinates and supercoordinates are functions of $\sigma$), then we
find the Hamiltonian of $1+1$ dimensionally reduced $10D$  
SYM theory in $A_0 = 0$ gauge,
with $x^9$ playing the role of the spatial component of the vector potential.
Thus the prescription of the previous section appears to be a particular
rule for how the large $N$ limit should be taken in the winding
configurations we have studied here. $P$ is taken to infinity first, and
then $M$ is taken to infinity.  The relation between the two
approaches is reminiscent of the Eguchi-Kawai\ref\eguchi{T.Eguchi, H.Kawai,
{\it Phys. Rev. Lett.} {\bf 48}, 1063, (1982).}
reduction of large $N$ gauge theory.  
It is clear once again that much of the physics 
of the matrix model is buried in the subtleties of
the large $N$ limit.  For multidimensional tori, the relationship
between the formalisms of this and the previous section is more obscure.

%
%
\newsec{Conclusions}

Although the  evidence we have given for the conjectured exact equivalence
between the large
$N$ limit of supersymmetric matrix quantum mechanics and uncompactified 11
dimensional M-theory is not definitive, it is quite substantial. The evidence
includes the following :

\item{1.} The matrix model has exact invariance under the Supergalilean group of
the \imf \ description of 11D Lorentz invariant theories.

\item{2.} Assuming the conventional duality between M-theory and IIA string
theory, the matrix model has  normalizable marginally bound states for any
value of $N$. These states have exactly the quantum numbers of the 11-D
supergraviton multiplet. Thus the spectrum of single particle states is exactly
that of M-theory.

\item{3.} As a consequence of supersymmetric nonrenormalization theorems,
asymptotic states of any number of non-interacting supergravitons exist. These
well separated particles propagate in a  Lorentz invariant manner in
11 dimensions.  They have the statistics properties of the SUGRA Fock space.

\item{4.} The matrix model exactly
reproduces
the correct long range interactions between supergravitons implied by
11-D supergravity, for zero longitudinal momentum exchange.
This one loop result could easily be ruined by higher loop effects
proportional to four powers of velocity. We believe that a highly nontrivial
supersymmetry theorem protects us against all higher loop corrections of this
kind.

\item{5.} By examining the pieces of the bound state wave function in
which two clusters of particles are well separated from each other,
a kind of mean field approximation, we
find that the longest range part of the wave
function grows with $N$ exactly as required by
the holographic principle. In particular the transverse density  never exceeds
one parton per Planck area.

\item{6.} The matrix model describes large classical membranes as required by
M-theory. The membrane world volume is a noncommutative space with a
fundamental unit of area analogous to the Planck area in phase space. These
basic quanta of area are the original \dzb \ from which the matrix model
was derived.  The tension of this matrix model membrane is precisely
the same as that of the M-theory membrane.

\item{7.} At the  classical level the matrix model realizes the full 11-D
Lorentz invariance in the large $N$ limit.

Of course  many unanswered questions remain.  Locality is extremely
puzzling in this system.  Longitudinal boost invariance, as we have
stressed earlier, is very mysterious.  Resolving this issue, perhaps by
understanding the intricate dynamics it seems to require, will be
crucial in deciding whether or not this conjecture is correct.

One way of understanding Lorentz invariance would be to
search for a covariant version of the matrix model in which the
idea of noncommutative geometry is extended to all of the membrane
coordinates.  An obvious idea is to consider functions of angular
momentum operators and try to exploit the connection between spin
networks and three dimensional diffeomorphisms.  Alternatively
one could systematically study
quantum corrections to the angular conditions.  

It is likely that more tests of the conjecture can be performed.
In particular it should be possible to examine
the large distance behavior of amplitudes with nonvanishing longitudinal
momentum transfer and to compare them with supergravity perturbation
theory.

It will be important to
to try to make precise the line of argument outlined in Section 9
that may lead to a proof of the conjecture.
The approaches to compactification discussed in Sections 9 and 10 should
be explored further.

If the conjecture is correct,
it would provide us with the first
well defined  nonperturbative formulation of a quantum theory
which includes gravitation. In principle, with a sufficiently big and fast
computer any scattering amplitude could be computed in the finite $N$ matrix
model with arbitrary precision. Numerical extrapolation to infinite $N$ is in
principle, if not in practice, possible. The situation is much like that in QCD
where the only known definition of the theory is in terms of a
conjectured limit  of lattice gauge theory. Although the practical utility
of the
lattice theory may be questioned, it is almost certain that an extrapolation to
the continuum limit exists. The existence of the lattice gauge Hamiltonian
formulation insures that the the theory is unitary and gauge invariant.

One can
envision the matrix model formulation of M-theory playing a similar role. It
would, among other things, insure that the rules of quantum mechanics are
consistent with gravitation. Given that the classical long distance equations
of 11-D supergravity have black hole solutions, a Hamiltonian formulation of
M-theory would, at last, lay to rest the claim that black holes lead to a
violation of quantum coherence.

\vfill\eject
\appendix{A}{}

In this appendix we will report on a preliminary investigation of the
threshold bound state wave function of $N$ zero branes in the large $N$
limit. In general, we may expect a finite probability for the $N$ brane
bound state to consist of $p$ clusters of $N_1 \ldots N_p$ branes
separated by large distances along one of the flat directions of the
potential.  We will try to take such configurations into account by 
writing a recursion relation relating the $N$ cluster to a $k$ and $N-k$
cluster.  This relation automatically incorporates multiple clusters
since the pair into which the original cluster is broken up will
themselves contain configurations in which they are split up into
further clusters.  There may however be multiple cluster configurations
which cannot be so easily identified as two such superclusters.  We will
ignore these for now, in order to get a first handle on the structure of
the wave function.

The configuration of a pair of widely separated clusters has a single
collective coordinate whose Lagrangian we have already written in our
investigation of supergraviton scattering.  The Lagrangian is 
\eqn\applag{L = \ha {k(N-k) \over N} v^2 + {k(N-k) \over r^7} v^4}
where $r$ is the distance between the clusters and $v$ is their relative
velocity.  By scaling, we can write the solution of this quantum
mechanics problem as $\phi ({{r (k[N-k])^{2/9}} \over N^{1/3}})$, where
$\phi$ is the threshold bound state wave function of the Lagrangian
\eqn\scaledlag{\ha v^2 + {v^4 \over r^7}~.}
This solution is valid when $r \gg l_p$.

We are now motivated to write the recursion relation
\eqn\recurs{\Psi_N = \Psi_N^{(c)} + \ha P \sum_{k = 1}^{N - 1} A_{N,k} \Psi_k
\Psi_{N-k} \phi ({{r (k[N-k])^{2/9}} \over N^{1/3}}) e^{- r Tr
W_k^{\dagger} W_k}}
Here we have chosen a gauge in order to make a block diagonal splitting
of our matrices.  $\Psi_j$ is the exact normalized threshold bound state wave
function for $j$ zero branes.  $W_k$ are the off diagonal $k \times N-k$
matrices which generate interactions between the two clusters.  $P$ is
the gauge invariant projection operator which rotates our gauge choice
among all gauge equivalent configurations.  The $A_{N.k}$ are
normalization factors, which in principle we would attempt to find by
solving the Schroedinger equation.  $\Psi_N^{(c)}  $ is the \lq\lq core
\rq\rq wave function, which describes configurations in which all of the
zero branes are at a distance less than or equal to $l_p$ from each
other.  We will describe some of its properties below.  In this regime,
the entire concept of distance breaks down, since the noncommuting parts
of the coordinates are as large as the commuting ones.

The interesting thing which is made clear by this ansatz, is that
the threshold bound state contains a host of internal distance scales,
which becomes a continuum as $N\rightarrow\infty$.  This suggests a
mechanism for obtaining scale invariant behavior for large $N$, as we
must if we are to recover longitudinal boost invariance.  Note that the
typical distance of cluster separation is largest as $N$ goes to
infinity when one of the clusters has only a finite number of partons.
These are the configurations which give the $N^{1/9}$ behavior discussed
in the text, which saturates the Bekenstein bound.  By the uncertainty
principle, these configurations have internal frequencies
of the bound state $\sim N^{- 2/9}$.  Although these go to zero as $N$
increases they are still infinitely higher than the energies of
supergraviton motions and interactions, which are of order $1/N$.  As in
perturbative string theory, we expect that this association of the large
distance part of the wave function with modes of very high frequency
will be crucial to a complete understanding of the apparent locality of
low energy physics.  

As we penetrate further in to the bound state, we encounter clusters of
larger and larger numbers of branes.  If we look for separated clusters
carrying finite fractions of the total longitudinal momentum, the
typical separation falls as $N$ increases.  
 Finally, we encounter the
core, $\Psi_N^{(c)}$, which we expect to dominate the ultimate short
distance and high energy behavior of the theory in noncompact eleven
dimensional spacetime.  

It is this core configuration to which the conventional methods of large
$N$ matrix models, which have so far made no appearance in our
discussion, apply.  Consider first gauge invariant Green's functions of
operators like $Tr X_i^{2k}$, where $i$ is one of the coordinate
directions.  We can construct a perturbation expansion of these Green's
functions by conventional functional integral methods.  When the time
separations of operators are all short compared to the eleven
dimensional Planck time, the terms in this expansion are well behaved.
We can try to resum them into a large $N$ series.  The perturbative
expansion parameter (the analog of $g^2_{YM}$ if we think of the theory
as dimensionally reduced Yang Mills theory) is ${R^3 \over l_p^6 E^3}$.
Thus, the planar Green's functions are functions of ${R^3 N \over l_p^6
E^3}$.

The perturbative expansion of course diverges term by term as
$E\rightarrow 0$.  If we imagine that, as suggested by our discussion
above, these Green's functions should be thought of as measuring
properties of the core wave function of the system, there is no physical
origin for such an infrared divergence.  If, as in higher dimensions,
the infrared cutoff is found already in the leading order of the $1\over
N$ expansion, then it must be of order $\omega_c \sim R l_p^{-2}
N^{1/3}$.  Note that this is much larger than any frequency encountered
in our exploration of the parts of the wave function with clusters
separated along a flat direction.  

Now let us apply this result to the computation of the infrared
divergent expectation values of single gauge invariant operators in the
core of the bound state wave function.  The idea is to evaluate the graphical
expansion of such an expression with an infrared cutoff and then insert
the above estimate for the cutoff to obtain the the correct large $N$
scaling of the object.  The combination of conventional large $N$
scaling and dimensional analysis then implies that planar graphs
dominate even though we are not taking the ``gauge coupling''
$~R^3 l_p^{-6}$, to zero as we approach the large $N$ limit.  Dimensional
analysis controls the otherwise unknown behavior of the higher order
corrections in this limit.   The results are
\eqn\lna{<{1\over N} tr X^{2k}> \sim N^{2k /3}}
\eqn\lnb{<{1\over N} tr [X^i , X^j ]^{2k}> \sim N^{4k /3}}
\eqn\lnc{<{1\over N} tr (\bar{\theta} [\gamma_i  X^i, \theta])^{2k}> \sim
N^{4k
/3}}
\eqn\lnd{ <{1\over N} tr \dot{X}^{2k}> \sim N^{4k /3}}
In the first of these expressions, $X$ refers to any component of the
transverse coordinates.  In the second the commutator refers to any pair
of the components.  The final expression, whose lowest order
perturbative formula has an ultraviolet divergence, is best derived by
combining \lnb , \lnc\ and the Schroedinger equation which says that the
threshold bound state has zero binding energy.  Note that these
expressions are independent of $R$, the compactification radius of the
eleventh dimension.  This follows from a cancellation between the $R$
dependence of the infrared cutoff, that of the effective coupling, and
that of the scaling factor which relates the variables $X$ to
conventionally normalized Yang Mills fields.

The first of these equations says that the typical eigenvalue of any one
coordinate matrix is of order $N^{1/3}$, much larger than the $N^{1/9}$
extension along the flat directions.  The second tells us that this
spectral weight lies mostly along the non flat directions.  In
conjunction the two equations can be read as a kind of \lq\lq
uncertainty principle of noncommutative geometry\rq\rq .  The typical
size of matrices is controlled by the size of their commutator.  The
final equation fits nicely with our estimate of the cutoff frequency.
The typical velocity is such that the transit time of a typical distance
\foot{in the space of eigenvalues, which in this noncommutative region,
is not to be confused with the classical geometrical distance between
\dzb \ .}is the inverse of the cutoff frequency.  It is clear that the
high velocities encountered in the core of the wave function could
invalidate our attempt to derive the matrix model by extrapolating from
weakly coupled string theory.  We must hope that the overall amplitude
for this part of the wave function vanishes in the large $N$ limit,
relative to the parts in which zero branes are separated along flat
directions. 

It is important to realize that these estimates do not apply along the
flat directions, but in the bulk of the $N^2$ dimensional configuration
space. In these directions, it does not make sense to multiply together
the \lq\lq sizes \rq\rq along different coordinate directions to make an
area since the different coordinates do not commute.  Thus, there is no
contradiction between the growth of the wave function in nonflat
directions and our argument that the size of the bound state in
conventional geometric terms, saturates the Bekenstein bound.

\appendix{B}{}
In this appendix we compute the matrix model membrane tension and 
show that it exactly agrees with the M-theory membrane tension.
The useful summary in \ref\dealwis{S. de Alwis, ``A Note on Brane
Tension and
M Theory,'' hep-th/9607011.} 
\def\ap{\alpha '}
\def\tpap{2 \pi \ap}
gives the tension of a D-p-brane $T_p$ in IIA
string theory or M-theory as
\eqn\tp{ T_p = {{(2\pi)^{1/2}} \over g_s} (2\pi)^{-p/2} ({1 \over \tpap})^
{{1+p} \over 2 }}
where $g_s$ is the fundamental string coupling and $1/\tpap$ is the
fundamental string tension.

The membrane tension $T_2$ is defined so that the mass ${\cal M}$ of a
stretched membrane of area $A$ is given by ${\cal M} = T_2 A$.
The mass squared for a light cone membrane with no transverse momentum
described by the map ${\cal X}^i(\sigma_1, \sigma_2),~ i=1 \ldots 9$ can be written
\eqn\memen{{\cal M}^2 = (2 \pi)^4 T_2^2 \int_0^{2 \pi} {d\sigma_1 \over
{2 \pi}}
\int_0^{2 \pi} {d\sigma_2 \over {2 \pi}} \sum_{i<j}
\lbrace {\cal X}^i , {\cal X}^j \rbrace^2 }
where the Poisson bracket of two functions $A(\sigma_1, \sigma_2),
B(\sigma_1, \sigma_2)$ is defined by
\eqn\pb{ \lbrace A,B \rbrace \equiv {\partial A \over \partial 
\sigma_1}{\partial B \over \partial 
\sigma_2} - {\partial B \over \partial 
\sigma_1}{\partial A \over \partial 
\sigma_2}.}

The coefficients in \memen\ are set by demanding that ${\cal M}^2$ for the
map ${\cal X}^8 = {\sigma_1 \over 2 \pi} L, ~{\cal X}^9
={\sigma_2 \over 2 \pi} L$
is given by ${\cal M}^2 = (T_2 L^2)^2$.

To understand the relation to the matrix model we
write, as in section 8, the map ${\cal X}^i$
as a fourier series
\eqn\fs{{\cal X}^i(\sigma_1, \sigma_2)=\sum_{n_1,n_2} x^i_{n_1 n_2}
e^{i(n_1\sigma_1+n_2\sigma_2)}.}
Then the corresponding matrices $X^i$ are given by 
\eqn\mats{X^i = \sum_{n_1,n_2} x^i_{n_1 n_2}U^{n_1}V^{n_2}}
where the matrices $U,V$ are elements of SU(N), have spectrum
$spec(U)=spec(V)=\{1,\omega, \omega^2 \ldots \omega^{N-1}\}$
and obey $UV=\omega VU$ where $\omega = \exp(2 \pi i/N)$.
In a specific basis, $U= diag(1,\omega, \omega^2 \ldots \omega^{N-1})$
and $V$ is a cyclic forward shift.

The scale of \mats\ is fixed since $spec(U)$ and $spec(V)$ 
go over as $N \rightarrow \infty$ to the unit circle $\exp(i \sigma)$.
Note that ${\cal X}^i$ is real so $X^i$ is hermitian.

The dynamics of the matrix model is governed by the Lagrangian
\eqn\lag{ {T_0 \over 2} Tr ( \sum_i\dot{X}^i \dot{X}^i 
+C\sum_{i<j}[X^i,X^j]^2) ~.}

The normalizations here are fixed by the requirement that \lag\
describe D-0-brane dynamics.  The first term, for diagonal matrices
describing D-0-brane motion, is just the nonrelativistic kinetic
energy ${m_0 \over 2} \sum_{a=1}^{N} v_a^2$ 
since the 0-brane mass $m_0 = T_0$. 
The rest energy of the system is just $N m_0$ which in the M-theory
interpretation is just $p_{11}$ so 
\eqn\pelev{p_{11} = N T_0.}  

The coefficient C is fixed by requiring that the small fluctuations
around diagonal matrices describe harmonic oscillators whose frequencies
are precisely the masses of the stretched strings connecting the
0-branes.  This ensures that \lag\ reproduces long range graviton
interactions correctly. Expanding \lag\ to quadratic order we find
that  $C=(1/\tpap)^2$.

The energy of a matrix membrane configuration with zero transverse
momentum is given by the commutator term in \lag .
We can evaluate this commutator in a semiclassical manner at large $N$
as in section 8
by introducing angular operators
$q$ with spectrum the interval $(0,2 \pi)$ and $p = {2 \pi \over N i} 
{\partial \over \partial q}$  with spectrum
the discretized interval $(0,2 \pi)$ so that $[p,q]= {2 \pi i \over
N}$.  The matrices  $U, V$ become $U=e^{ip}, V=e^{iq}$.
By Baker-Campbell-Hausdorff 
we see $UV=\omega VU$.
The
formal $\hbar$ in this algebra is given by $\hbar = {2 \pi \over N}$.
Semiclassically we have
\eqn\semicl{\eqalign{[X,Y] &\rightarrow i \hbar \{ {\cal X}, {\cal Y} \} \cr
Tr &\rightarrow \int_0^{2 \pi}\int_0^{2 \pi} {dp dq \over 2 \pi \hbar}
.}}
So we get 
\eqn\semicom{Tr[X^i, X^j]^2 \rightarrow - { (2 \pi)^2 \over N}
\int_0^{2 \pi} {d\sigma_1 \over {2 \pi}}
\int_0^{2 \pi} {d\sigma_2 \over {2 \pi}} 
\lbrace {\cal X}^i , {\cal X}^j \rbrace^2 . }

This commutator can also be evaluated for a given finite $N$ matrix 
configuration explicitly with results that agree with \semicom\ as 
$N \rightarrow \infty$.

Now we can perform the check.  The value of the matrix model Hamiltonian
on a configuration with no transverse momentum is
\eqn\h{H = {T_0 \over 2} ({1 \over \tpap})^2{(2 \pi)^2 \over  N} 
\int_0^{2 \pi} {d\sigma_1 \over {2 \pi}}
\int_0^{2 \pi} {d\sigma_2 \over {2 \pi}} 
\sum_{I<j}\lbrace {\cal X}^i , {\cal X}^j \rbrace^2  .}
The conjecture interprets the matrix model Hamiltonian $H$ as the 
infinite momentum frame energy $\sqrt{p_{11}^2 + {\cal M}^2} - p_{11}
\simeq  {{\cal M}^2 \over 2 p_{11}}$. 
So the matrix membrane mass squared is ${\cal M}^2_{\rm mat} =
2 p_{11}H.$ Using \pelev\ we find
\eqn\matmem{{{\cal M}^2_{\rm mat}} =  {T_0^2 } ({1 \over \tpap})^2
{(2 \pi)^2} 
\int_0^{2 \pi} {d\sigma_1 \over {2 \pi}}
\int_0^{2 \pi} {d\sigma_2 \over {2 \pi}} \sum_{i<j}
\lbrace {\cal X}^i , {\cal X}^j \rbrace^2  .}
{}From \memen\ we can now read off the matrix model membrane tension
as 
\eqn\mattens{(T_2^{~\rm mat})^2 =  {T_0^2 } ({1 \over \tpap})^2
{1 \over (2 \pi)^2} .}
So we can write
\eqn\ratio{\eqalign{{T_2^2 \over (T_2^{~\rm mat})^2} &= (2\pi)^2 (\tpap)^2 
({T_2 \over T_0})^2 \cr
&=(2 \pi)^2 ({1 \over 2 \pi})^2 \cr
&= 1 .}}

So the M-theory and matrix model membrane tensions exactly agree.

\vfill\eject
\centerline{\bf Acknowledgments}

This work crystallized at the Strings 96 Conference in Santa Barbara, and a
very preliminary version of it was presented there.  We would like to
thank the organizers and participants of this conference for providing
us with a stimulating venue for the production of exciting physics.  We
would particularly like to thank M. Green for pointing out the relation
to previous work on supermembranes, and B. de Wit and I. Bars for
discussing some of their earlier work on this topic with us.  
We would also like to thank C. Bachas and M. Dine for 
conversations on nonrenormalization
theorems, M. Dine for discussing his work with us,
C. Thorn for discussions about light cone
string theory, N. Seiberg for discussions about supersymmetry, 
and M. Berkooz, M. Douglas and N. Seiberg for 
discussions about membrane tension.
T.B.,
W.F., and S.H.S. would like to thank the Stanford Physics Department for
its hospitality while some of this work was carried out.
The work of T.B. and S.H.S. was
supported in part by the Department of Energy under Grant No. DE - FG02
- 96ER40959 and that  of W.F. was supported in part by the Robert A. Welch
Foundation and by NSF Grant PHY-9511632.
L.S. acknowledges the support of the NSF under Grant No. PHY - 9219345.

\listrefs
\end